\documentclass[aps,prl,twocolumn,showpacs]{revtex4-1}
\usepackage{amsmath}
\usepackage{amssymb}
\usepackage{graphicx}
\usepackage{breakurl}
\usepackage{siunitx}
\usepackage[T1]{fontenc}
\usepackage[latin9]{inputenc}
\usepackage{color}
\usepackage{amstext}
\usepackage{graphicx}
\usepackage[normalem]{ulem}
\usepackage[unicode=true,pdfusetitle,pdfborder={0 0 1}]{hyperref}
\hypersetup{colorlinks,
linkcolor=blue,          
citecolor=blue,        
filecolor=blue,      
urlcolor=blue           
}
\usepackage{upgreek}
\usepackage{mathrsfs}
\usepackage{amsbsy}
\usepackage{epstopdf}
\usepackage{esint}

\makeatletter
\@ifundefined{textcolor}{}
{
 \definecolor{BLACK}{gray}{0}
 \definecolor{WHITE}{gray}{1}
 \definecolor{RED}{rgb}{1,0,0}
 \definecolor{GREEN}{rgb}{0,1,0}
 \definecolor{BLUE}{rgb}{0,0,1}
 \definecolor{CYAN}{cmyk}{1,0,0,0}
 \definecolor{MAGENTA}{cmyk}{0,1,0,0}
 \definecolor{YELLOW}{cmyk}{0,0,1,0}
 }
\makeatother

\newcommand{\ket}[1]{| #1 \rangle} 

\graphicspath{{../figures/}}

\begin{document}

\title{Probe of Three-Dimensional Chiral Topological Insulators in an
Optical Lattice}
\author{S.-T. Wang, D.-L. Deng, and L.-M. Duan}
\affiliation{Department of Physics, University of Michigan, Ann Arbor, Michigan 48109, USA}
\affiliation{Center for Quantum Information, IIIS, Tsinghua University, Beijing 100084,
PR China}

\begin{abstract}
We propose a feasible experimental scheme to realize a three-dimensional chiral
topological insulator with cold fermionic atoms in an optical lattice, which
is characterized by an integer topological invariant distinct from the conventional $\mathbb{Z}_{2}$  topological insulators and has a remarkable macroscopic zero-energy flat band. To probe its property, we show that its characteristic surface states --- the Dirac cones --- can be probed through time-of-flight imaging or Bragg spectroscopy and the flat band can be detected via measurement of the atomic density profile in a weak global trap. The realization of this novel topological phase with a flat band in an optical lattice will provide a unique experimental platform to study the interplay between
interaction and topology and open new avenues for application of topological states.
\end{abstract}

\maketitle

The exploration of topological phases of matter has become a major theme at
the frontiers of condensed matter physics since the discovery of
topological insulators (TIs) \cite{hasan2010colloquium, *Qi:2011wt,*moore2010birth}. The TIs are band insulators with peculiar topological properties that are protected by time reversal symmetry. A recent remarkable theoretical advance is the
finding that there are various other kinds of topological phases of free
fermions apart from the conventional TIs, which can be classified by a
periodic table according to system symmetry and dimensionality \cite%
{Schnyder:2008ez,*kitaev2009periodic}. An important question then is whether
the new topological phases predicted by the periodic table can be physically
realized. Several model Hamiltonians have been proposed to have the
predicted topological phases as their ground states \cite{Ryu:2010ko,neupert2012noncommutative,qi2009time,Schnyder:2009he,moore2008topological,*Deng:2013fe}. However, these model Hamiltonians typically require complicated spin-orbital couplings that are hard to be realized in real materials. Implementations of these model Hamiltonians still remain very challenging for experiments.

In this Letter, we propose an experimental scheme to realize a
three-dimensional (3D) chiral TI with cold fermionic atoms in an optical
lattice. The chiral TI is protected by the chiral symmetry, also known as
the sublattice symmetry \cite{Ryu:2010ko,Hosur:2010ie,essin2012antiferromagnetic}. Unlike the conventional TIs protected by the time reversal symmetry, which is characterized by a $\mathbb{Z}_{2}$ topological
invariant, the chiral TI is characterized by a topological invariant taking
arbitrary integer values \cite{Schnyder:2008ez,*kitaev2009periodic}. By
controlling the spin-orbital coupling of cold fermionic atoms in a tilted
optical lattice based on the Raman-assisted hopping \cite{jaksch2003creation,Miyake:2013jw,Aidelsburger:2013ew}, we realize a tight-binding model Hamiltonian first proposed in Ref.\ \cite{neupert2012noncommutative}, which supports a chiral TI with a zero-energy flat band. In such a flat band, the kinetic energy is suppressed and the atomic interaction, which can be tuned by the Feshbach resonance technique \cite%
{chin2010feshbach}, will lead to a novel nonperturbative effect.  In a cold atom experiment, flat bands have been studied in a 2D frustrated Kagome lattice \cite{jo2012ultracold}. Inspired by the discovery of the fractional quantum Hall effect in a topologically
nontrivial flat-band Landau level, one expects that the atomic interaction
in a flat-band TI may lead to exciting new physics \cite%
{tang2011high,*neupert2011fractional,*sun2011nearly}. To probe the properties of the chiral TI in our proposed realization, we show that topological phase transition and the characteristic surface states of the TIs, the Dirac cones, can both be detected by mapping out the Fermi surface structure through time-of-flight imaging \cite{Spielman2007Mott, *Kashurnikov2002Revealing, kohl2005fermionic} or Bragg spectroscopy \cite{stamper1999excitation}. Furthermore, we show that the flat band can be verified by measurement of the atomic density profile under a weak global harmonic trap \cite{zhu2007simulation,Schneider05122008}.

We consider realization of the following tight-binding model Hamiltonian in
the momentum space \cite{neupert2012noncommutative}%
\begin{equation}
\mathcal{H}(\mathbf{k})=\left(
\begin{array}{ccc}
0 & 0 & q_{1}-iq_{2} \\
0 & 0 & q_{3}-iq_{0} \\
q_{1}+iq_{2} & q_{3}+iq_{0} & 0%
\end{array}%
\right) ,  \label{Eqn:Ham}
\end{equation}%
with $q_{0}=2t\left( h+\cos k_{x}a+\cos k_{y}a+\cos k_{z}a\right)
$, $q_{1}=2t\sin k_{x}a$, $q_{2}=2t\sin k_{y}a$, $q_{3}=2t\sin k_{z}a$, where $%
\mathbf{k=}\left( k_{x},k_{y},k_{z}\right) $ denotes the momentum, $a$ is
the lattice constant, $t$ is the hopping energy, and $h$ is a dimensionless control parameter. This model Hamiltonian has a chiral symmetry represented by $S\mathcal{H}(\mathbf{k})S^{-1}=-\mathcal{H}(\mathbf{k})$ with the unitary matrix $S \equiv $diag$(1,1,-1)$. It has three bands, with a flat middle band exactly at zero energy protected by the chiral symmetry. The other two bands have energies $E_{\pm }(\mathbf{k})=\pm
2t[\sin ^{2}\left( k_{x}a\right) +\sin ^{2}\left( k_{y}a\right) +\sin
^{2}\left( k_{z}a\right) +(\cos k_{x}a+\cos k_{y}a+\cos k_{z}a+h)^{2}]^{1/2}$%
. The topological index for this model can be characterized by the integral
\cite{neupert2012noncommutative,deng2013systematic}
\begin{equation}
\Gamma =\frac{1}{12\pi ^{2}}\int_{\text{BZ}}d\mathbf{k}\;\epsilon ^{\alpha
\beta \gamma \rho }\epsilon ^{\mu \nu \tau }\frac{1}{E_{+}^{4}}q_{\alpha
}\partial _{\mu }q_{\beta }\partial _{\nu }q_{\gamma }\partial _{\tau
}q_{\rho },
\label{Eqn:Index}
\end{equation}%
where $\epsilon $ is the Levi-Civita symbol with $\left( \alpha ,\beta
,\gamma ,\rho \right) $ and $\left( \mu ,\nu ,\tau \right) $ taking values
respectively from $\left\{ 0,1,2,3\right\} $ and $\left\{ k_{x},k_{y},k_{z}\right\} $.

To realize the model Hamiltonian \eqref{Eqn:Ham}, we consider interaction-free
fermionic atoms in an optical lattice and choose three internal atomic
states in the ground state manifold to carry three spin states $\ket{1},\ket{2},\ket{3}$. The other levels in the ground state manifold are irrelevant as they are initially depopulated by the optical pumping and transitions to these levels are forbidden during Raman-assisted atomic hopping because of a large energy detuning. The Hamiltonian \eqref{Eqn:Ham}, expressed in real space, has the following form
\begin{align}
H&= t \sum_{\mathbf{r}}\left[ \left( 2ihc_{3,\mathbf{r}%
}^{\dag }c_{2,\mathbf{r}}+\text{H.c.}\right) +H_{\mathbf{rx}}+H_{\mathbf{ry}%
}+H_{\mathbf{rz}}\right] \notag, \\
H_{\mathbf{rx}}& =ic_{3,\mathbf{r-x}}^{\dag }(c_{1,\mathbf{r}}+c_{2,\mathbf{%
r}})-ic_{3,\mathbf{r+x}}^{\dag }(c_{1,\mathbf{r}}-c_{2,\mathbf{r}})+%
\text{H.c.,}  \notag \\
H_{\mathbf{ry}}& =-c_{3,\mathbf{r-y}}^{\dag }(c_{1,\mathbf{r}}-ic_{2,\mathbf{%
r}})+c_{3,\mathbf{r+y}}^{\dag }(c_{1,\mathbf{r}}+ic_{2,\mathbf{r}})+%
\text{H.c.,}  \notag \\
H_{\mathbf{rz}}& =2ic_{3,\mathbf{r-z}}^{\dag }c_{2,\mathbf{r}}+\text{H.c.},
\label{Eqn:Ham2}
\end{align}%
where $\left( \mathbf{x,y,z}\right) $ represents a unit vector along the $\left( x,y,z\right) $-direction of a cubic lattice, and $c_{j,\mathbf{r}}$ $\left( j=1,2,3\right) $\ denotes the
annihilation operator of the fermionic mode at the lattice site $\mathbf{r}$
with the spin state $\ket{j}$. To implement this Hamiltonian, the major
difficulty is to realize the spin-transferring hopping terms $H_{\mathbf{rx}%
},H_{\mathbf{ry}},H_{\mathbf{rz}}$ along each direction \cite{OpticalLatCTI:Sup}. The hopping terms and the associated spin transformation can be visualized diagrammatically as
\vspace{-.1cm}
\begin{align}
\hspace{-0.2cm}x& \text{-}\text{direction:}\,\ket{3}\overset{i\sqrt{2}}{\curvearrowleft}%
\ket{1_x}\overset{\times }{\curvearrowright }\;+\;\overset{\times }{%
\curvearrowleft }\ket{2_x}\overset{-i\sqrt{2}}{\curvearrowright }\ket{3}\;+\;\text{%
H.c.} \notag \\
\hspace{-0.2cm}y& \text{-}\text{direction:}\,\ket{3}\overset{-\sqrt{2}}{\curvearrowleft }%
\ket{1_y}\overset{\times }{\curvearrowright }\;+\;\overset{\times }{%
\curvearrowleft }\ket{2_y}\overset{\sqrt{2}}{\curvearrowright }\ket{3}\;+\;\text{%
H.c.} \notag \\
\hspace{-0.2cm}z& \text{-}\text{direction:}\,\ket{3}\overset{2i}{\curvearrowleft }%
\ket{2}\overset{\times }{\curvearrowright }\;+\;\text{H.c.}
\label{Eqn:Hopping}
\end{align}
where $\overset{\times }{\curvearrowright }$ indicates that hopping is
forbidden along that direction, and $\ket{1_x}=\left(\ket{1}+\ket{2}\right)/\sqrt{2},\ket{2_x}=(\ket{1}-\ket{2})/\sqrt{2},\ket{1_y}=(\ket{1}-i\ket{2})/\sqrt{2},\ket{2_y}=(\ket{1}+i\ket{2})/\sqrt{2}$ are superpositions of the original spin-basis vectors $\ket{1},%
\ket{2},\ket{3}$.

We use Raman-assisted tunneling to achieve the spin-transferring hopping
terms depicted in Eq.\ \eqref{Eqn:Hopping}. Note that the parity (left-right) symmetry is
explicitly broken by these hopping terms. To break the parity symmetry, we
assume the optical lattice is tilted with a homogeneous energy gradient
along the $x$-,$y$-,$z$-directions. This can be achieved, for
instance, through the natural gravitational field, the magnetic field gradient,
or the gradient of a dc- or ac-Stark shift \cite{jaksch2003creation,Miyake:2013jw,Aidelsburger:2013ew}. Raman-assisted hopping in a tilted optical lattice has been
demonstrated in recent experiments \cite{Miyake:2013jw,Aidelsburger:2013ew}. In our scheme, we require a different
linear energy shift per site $\Delta _{x,y,z}$ along the $x$-,$y$-,$z$-directions. In particular, we take $%
\Delta _{z}\approx 1.5\Delta _{y}\approx 3\Delta _{x}$ with the energy
difference lower bounded by $\Delta _{x}$, and assume the natural tunneling
rate $t_{0}\ll \Delta _{x}$ so that the hopping probability $\left(
t_{0}/\Delta _{x}\right) ^{2}$ induced by the natural tunneling is
negligible in this tilted lattice.

\begin{figure}[b]
\hspace{.8cm}(a) laser configuration\\
\includegraphics[width=0.4\textwidth]{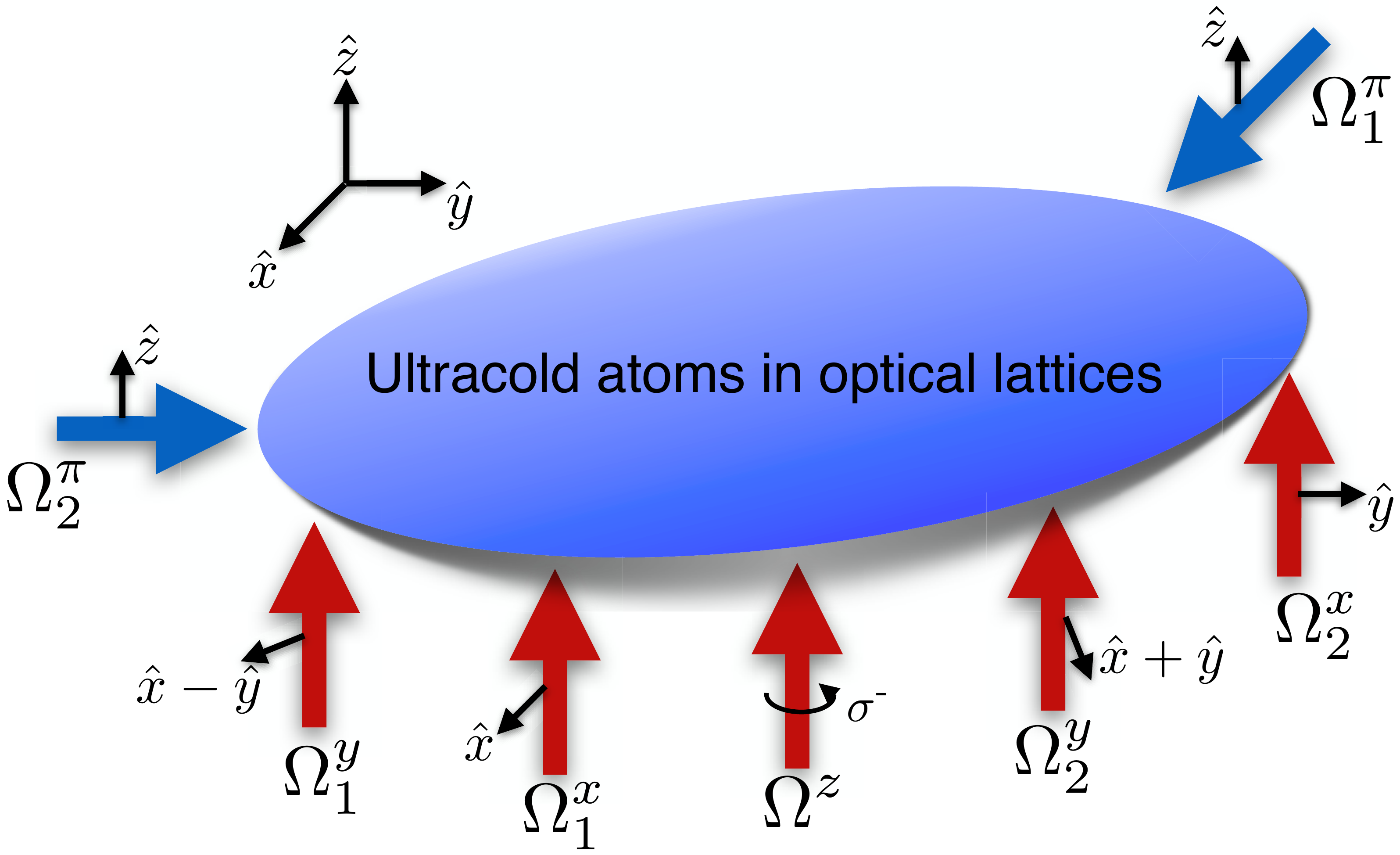}\\
\hspace{-.8cm}(b) tilted optical lattice \hspace{1.6cm} (c) $x$-direction\\
\hspace{.1cm}\includegraphics[width=0.18\textwidth]{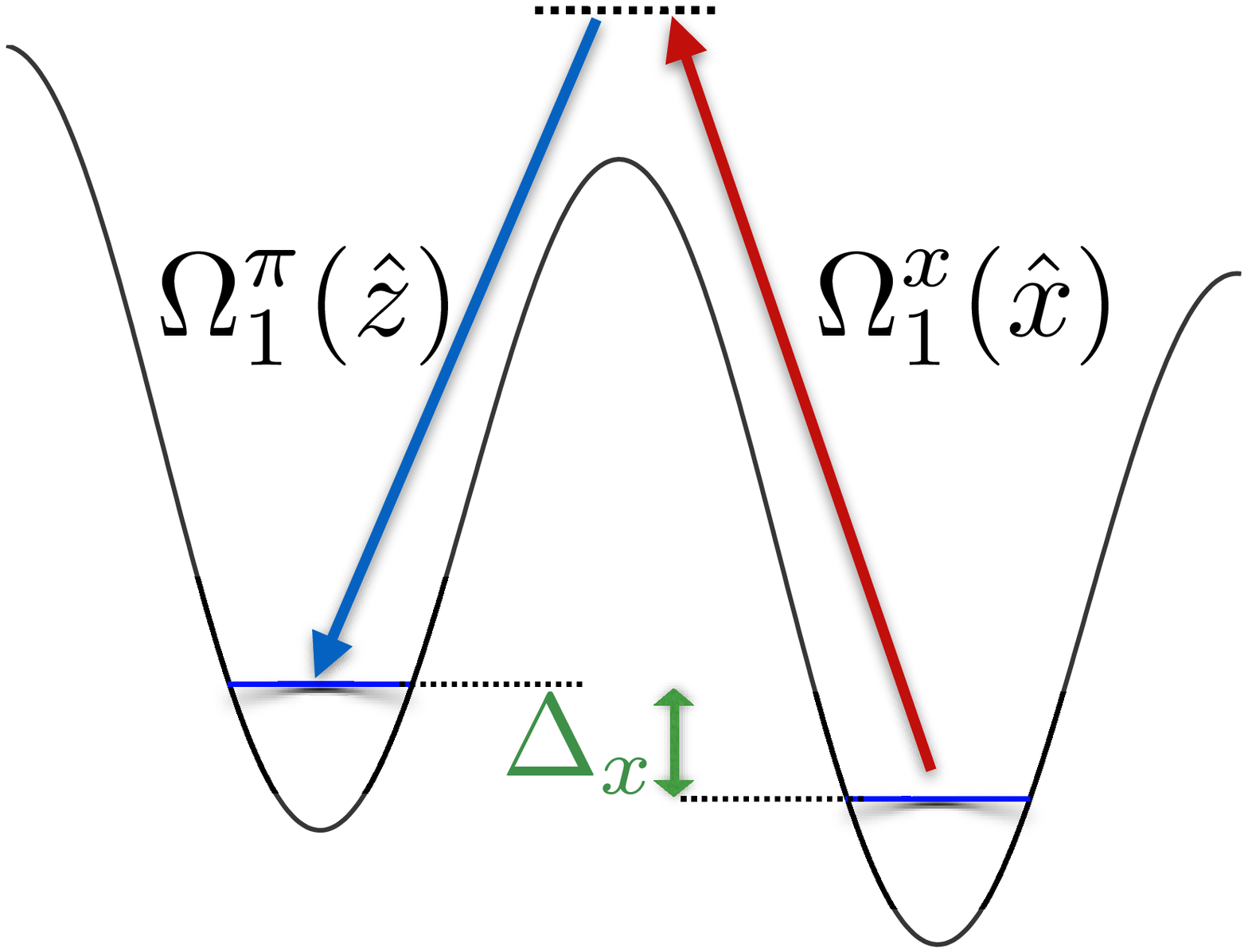}
\hspace{.55cm}
\includegraphics[width=0.22\textwidth]{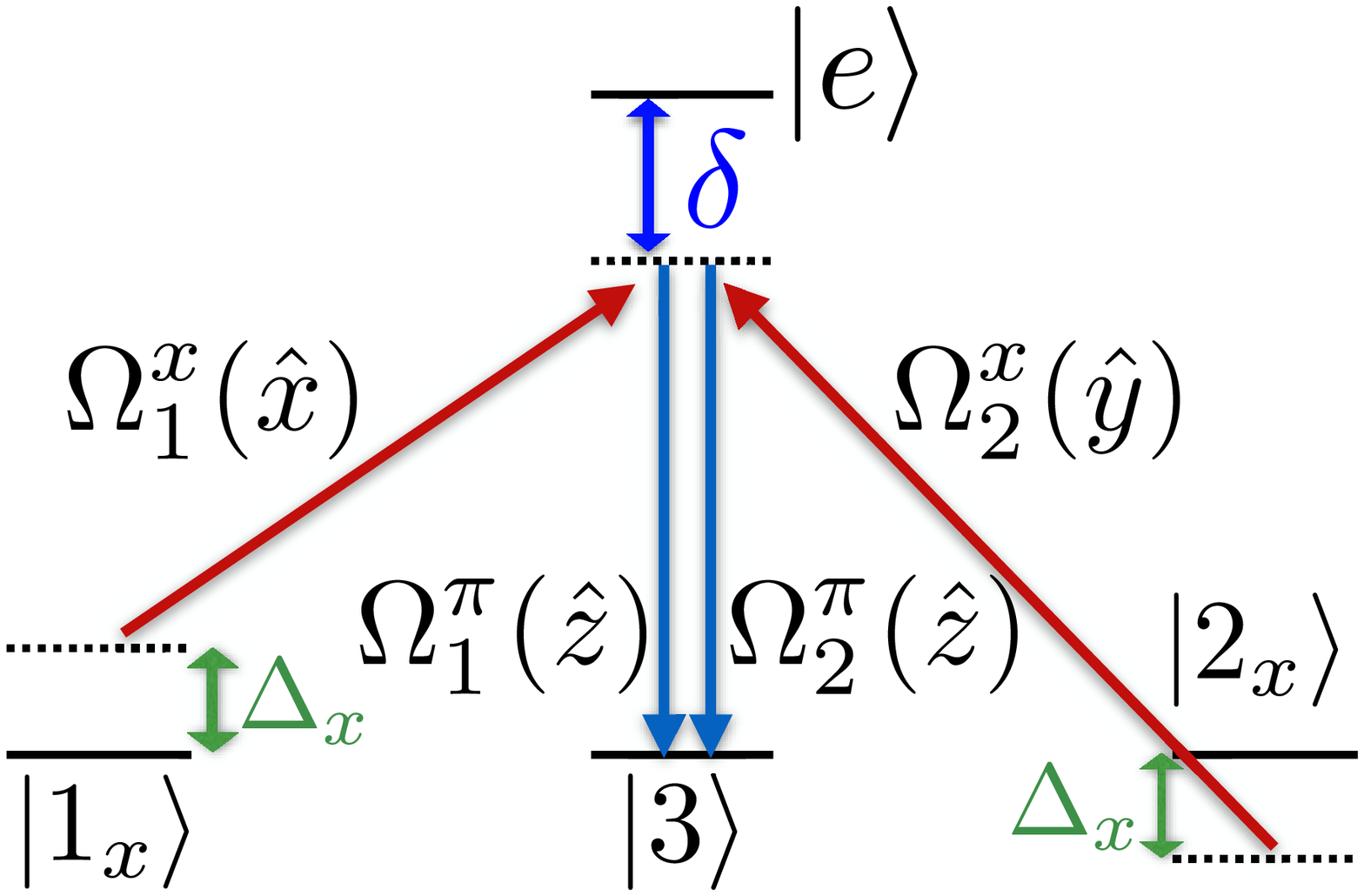}\\
(d) $y$-direction \hspace{2.1cm} (e) $z$-direction\\
\includegraphics[width=0.21\textwidth]{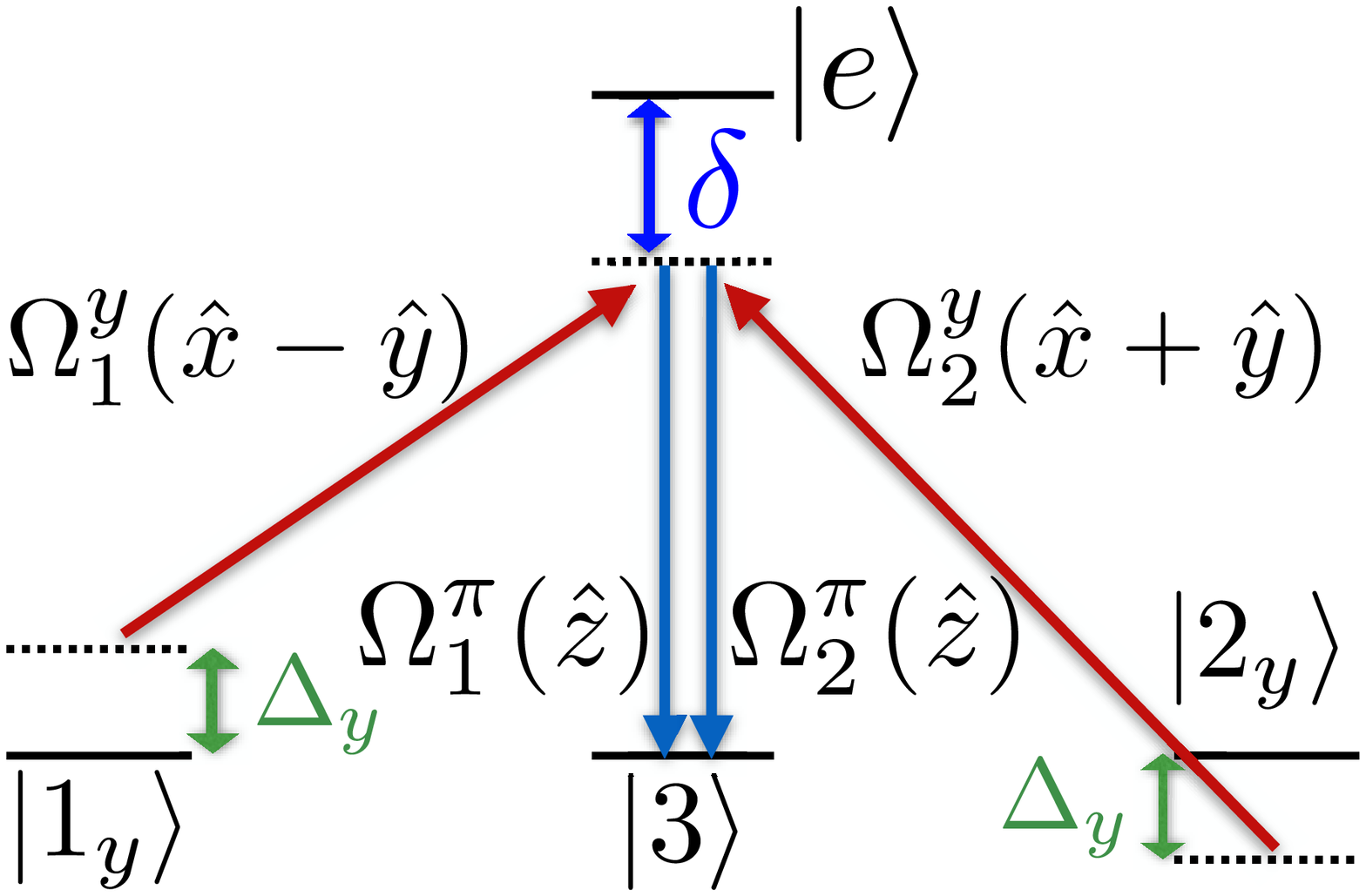}
\hspace{.3cm}
\includegraphics[width=0.21\textwidth]{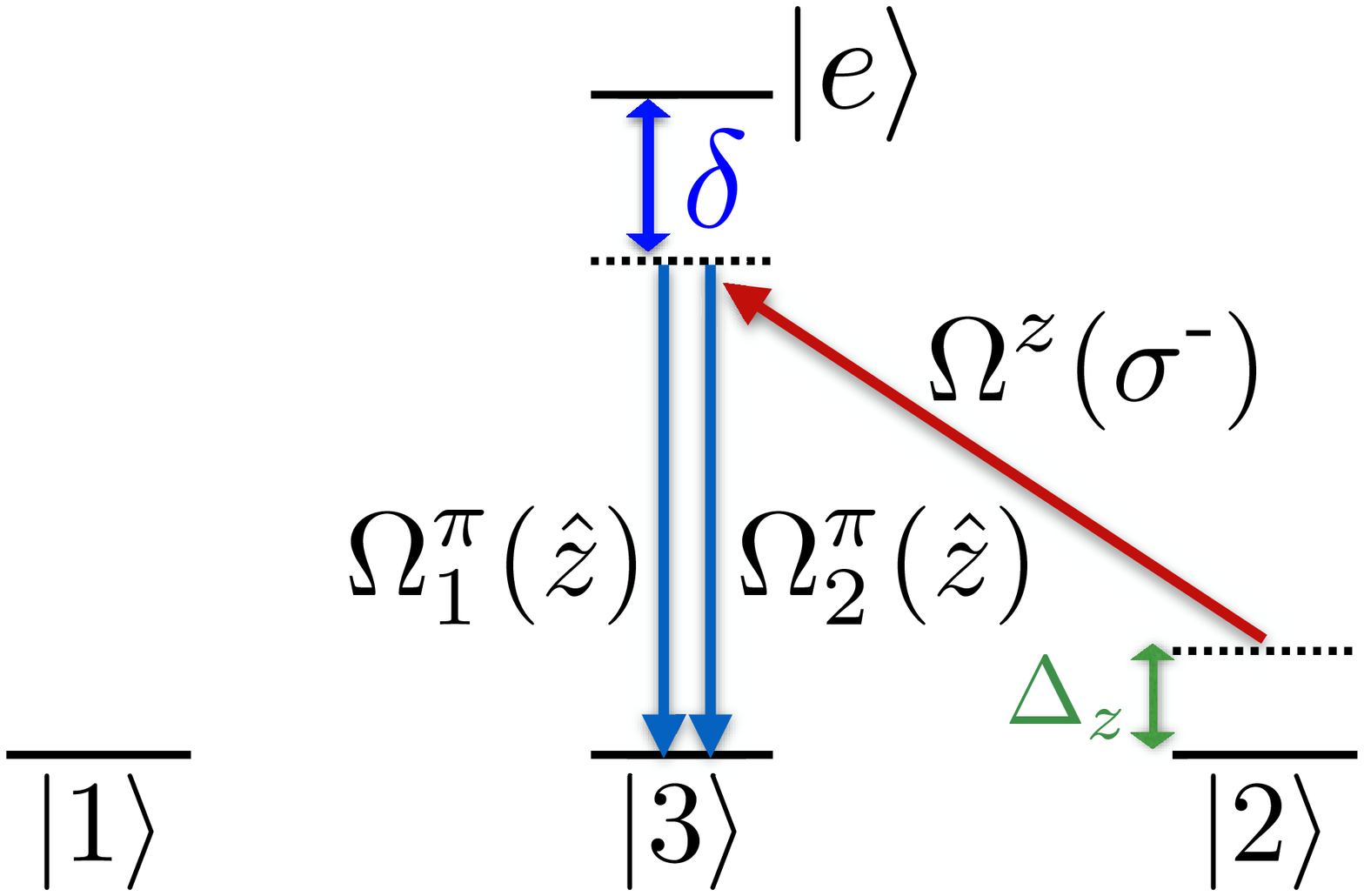}
\caption{Schematics of the laser configuration to realize the Hamiltonian in Eq.\ \eqref{Eqn:Ham2}. Panel (a) shows the propagation direction (big arrows) and the polarization (small arrows) of each laser beam. (b) A linear tilt $\Delta_{x,y,z}$ per site in the lattice along each direction. The detuning in each direction matches the frequency offset of the corresponding Raman beams, which are shown in panels (c), (d), and (e). Polarizations of each beam are shown in brackets. Rabi frequencies for each beam are: $\Omega _{1}^{\pi }=\Omega _{0}e^{ikx}, \Omega _{2}^{\pi}=\Omega _{0}e^{iky}, \Omega _{1}^{x}=i\sqrt{2}\Omega _{0}e^{ikz}, \Omega _{2}^{x}=-i\sqrt{2}\Omega _{0}e^{ikz}, \Omega _{1}^{y} =-\sqrt{2}\Omega _{0}e^{ikz}, \Omega _{2}^{y}=\sqrt{2}\Omega _{0}e^{ikz}, \Omega ^{z} =2i\Omega _{0}e^{ikz}$. \cite{OpticalLatCTI:Sup}}
\label{Fig:OpticalLattice}
\end{figure}

To realize the hopping terms in Eq.\ \eqref{Eqn:Hopping}, we apply
two-photon Raman transitions with the configuration (polarization and
propagating direction) of the laser beams depicted in Fig.\ \ref%
{Fig:OpticalLattice} \cite{OpticalLatCTI:Sup}. The internal states $\ket{1},\ket{3}, \ket{2}$ differ in the magnetic quantum number $m$ by one successively so that the atomic addressing can be achieved using polarization selection. The $\pi$-polarized lights consist of two laser beams $\Omega _{1}^{\pi }=\Omega _{0}e^{ikx}$ and $\Omega _{2}^{\pi }=\Omega
_{0}e^{iky}$, propagating along the $x$ and $y$ directions respectively,
where $k$ is the magnitude of the laser wave vector. The other five beams $%
\Omega _{1,2}^{x,y,z}$ are all propagating along the $z$ direction and the
polarizations are shown in Fig.\ \ref{Fig:OpticalLattice}. The Rabi
frequencies $\Omega _{1,2}^{x,y,z}$, expressed in terms of the unit $\Omega
_{0}$, are given in the caption of Fig.\ \ref{Fig:OpticalLattice}  to produce the required phase and amplitude relations of the hopping terms in Eq.\ \eqref{Eqn:Hopping}.
Between the sites $\mathbf{r}$ and $\mathbf{r+m}$, the Raman-assisted
hopping rate is given by
\begin{equation*}
t_{\mathbf{r,m}}=\dfrac{\Omega _{\beta \mathbf{m}}^{\ast }\Omega _{\alpha
\mathbf{m}}}{\delta }\int d^{3}\mathbf{r}^{\prime }w^{\ast }(\mathbf{r}%
^{\prime }-\mathbf{r-m})e^{i\delta \mathbf{k}\cdot \mathbf{r}^{\prime }}w(%
\mathbf{r}^{\prime }-\mathbf{r}),
\end{equation*}%
where $\delta $ is a large single-photon detuning to the excited state, $w(%
\mathbf{r}^{\prime }-\mathbf{r})$ is the Wannier-(Stark) function at the site $\mathbf{r}$ \cite{OpticalLatCTI:footnote1}, and $\delta \mathbf{k=k}_{\alpha}-\mathbf{k}_{\beta}$ is the momentum difference between the relevant Raman beams with the corresponding single-photon Rabi frequencies $\Omega _{\alpha \mathbf{m}}$ and $\Omega_{\beta \mathbf{m}}$. Because of the fast decay of the Wannier function, we consider only the nearest-neighbor Raman-assisted hopping with $\mathbf{m=} \pm \mathbf{x},\pm \mathbf{y},\pm \mathbf{z}$. When $\delta \mathbf{k=0}$, we have $t_{\mathbf{r,m}}=0$ for any $\mathbf{m\neq 0}$ terms because of the orthogonality of Wannier functions. Let us take one of the tunneling terms along the $x$ direction $\ket{3}\overset{i\sqrt{2}}{\curvearrowleft }\ket{1_x}$ as an example to explain the Raman-assisted hopping rate. The relevant Raman pair is $\Omega _{1}^{x}=i\sqrt{2}\Omega _{0}e^{ikz}$ and $\Omega _{1}^{\pi}=\Omega _{0}e^{ikx}$ in Fig.\ \ref{Fig:OpticalLattice}, so $\Omega _{\alpha \mathbf{m} }=i\sqrt{2}\Omega _{0}$ and $\Omega _{\beta \mathbf{m}}=\Omega _{0}$.
The laser beam $\Omega _{1}^{x}$ has two frequency components, generated, e.g., by an electric optical modulator (EOM), which are resonant with the levels $\ket{1},\ket{2}$ respectively so that in the rotating frame the levels $\ket{1}$ and $\ket{2}$ are degenerate in energy. The beam $\Omega _{1}^{x}$ is polarized along the $x$ direction, so, together with $%
\Omega _{1}^{\pi }$, it couples the state $\ket{1_x}$ to the state $\ket{3}$ through the two-photon transition. The two-photon detuning $\Delta _{x}$ is in resonance with the potential gradient along the $x$ direction so that the beams only induce the
nearest-neighbor hopping from $\mathbf{r}$ to $\mathbf{r-x}$. Using
factorization of the Wannier function $w(\mathbf{r}^{\prime
})=w(x^{\prime })w(y^{\prime })w(z^{\prime })$ in a cubic lattice, we find
the hopping rate $t_{\mathbf{r,-  x}}=i\sqrt{2}\beta \Omega _{\mathbf{R}}e^{i\delta
\mathbf{k}\cdot \mathbf{r}}$, where $\Omega _{\mathbf{R}}\equiv \left\vert
\Omega _{0}\right\vert ^{2}/\delta $ and $\beta \equiv \int dxw^{\ast
}(x+a)e^{-ikx}w(x)\int dyw^{\ast }(y)w(y)\int dzw^{\ast }(z)e^{ikz}w(z)$.
For this hopping term, $\delta \mathbf{k}=(-k,0,k)$. Actually, for the beams
shown in Fig.\ \ref{Fig:OpticalLattice}, any nonzero $\delta \mathbf{k}$ has
the form $(\pm k,0,\mp k)$ or $(0,\pm k,\mp k)$, so the site dependent phase
term can always be reduced to $e^{i\delta \mathbf{k}\cdot \mathbf{r}}=1$ if
we take the lattice constant $a$ to satisfy the condition $ka=2\pi $ by
adjusting the interfering angle of the lattice beams. Under this condition,
all the hopping terms in Eq.\ \eqref{Eqn:Hopping} are obtained through the laser beams shown in Fig.\ \ref{Fig:OpticalLattice} with the hopping rate $t=\beta \Omega _{\mathbf{R}}$ \cite{OpticalLatCTI:Sup}. The on-site spin transferring term $hc_{3,\mathbf{r}}^{\dag }c_{2,\mathbf{r}}$ can be achieved through
application of a simple radio-frequency (rf) field (or another
copropagating Raman beam). The Raman beams $\Omega _{1,2}^{x,y,z}$ and $%
\Omega _{1,2}^{\pi}$ may also induce some on-site spin transferring terms,
which can be similarly compensated (canceled) with additional rf fields.

Although the laser configuration illustrated in Fig.\ \ref{Fig:OpticalLattice} involves several beams, all of them can be drawn from the same laser, with the small relative
frequency shift induced by an acoustic optical modulator (AOM) or EOM. The
absolute frequencies of these beams and their fluctuations are not important
as long as we can lock the relative frequency differences, which can be well
controlled with the driving rf fields of the AOMs and EOMs. To show that
the proposed scheme is feasible with current technology, we give a
parameter estimation for typical experiments. For instance, with $^{40}$K
atoms of mass $m$ in an optical lattice with the lattice constant $a=2\pi
/k=764$ nm \cite{liu2013realization,Wang2012Spin}, gravity induces a potential gradient (per site) $\Delta =mga/\hbar\approx 2\pi \times 0.75\,$kHz. Gravity provides the gradients for free along three directions with an appropriate choice of the relative axes of the frame to satisfy $\Delta _{x}:\Delta _{y}:\Delta _{z}=1:2:3$ and $\Delta
=\sqrt{\Delta _{x}^{2}+\Delta _{y}^{2}+\Delta _{z}^{2}}$. We then have $%
\Delta _{x}\approx 2\pi \times 200\,$Hz. For a lattice with depth $V_{0}\approx 2.3E_{r}$,
where $E_{r}=\hbar ^{2}k^{2}/2m$ is the recoil energy, the overlap ratio $%
\beta \approx 0.34$ and the natural tunneling rate $t_{0}/\hbar \approx 2\pi \times 50\,$Hz \cite{OpticalLatCTI:Sup}. For Raman beams with $\Omega _{0}/2\pi \approx 15\,$MHz and the single-photon detuning $\delta /2\pi \approx 1.7\,$THz \cite{liu2013realization}, we have $\Omega _{\mathbf{R}}=\left\vert \Omega _{0}\right\vert ^{2}/\delta \approx 2\pi \times 120\,$Hz and the Raman-assisted hopping rate $t/\hbar \approx 2\pi \times 40\,$Hz. Apparently, the undesired off-resonant hopping
probabilities, upper bounded by $t^{2}/\Delta _{x}^{2}$ or $t_{0}^{2}/\Delta
_{x}^{2}$, are less than $6\%$ and the effective spontaneous emission rate,
estimated by $|\Omega _{0}/\delta |^{2}\Gamma _{s}$ ($\Gamma _{s}\approx 2\pi \times 6\,$MHz is the decay rate of the excited state), is negligible during the
experimental time of the order of $10/t$.

\begin{figure}[b]
\includegraphics[trim=0.3cm 0.2cm 0cm 0.4cm, clip,width=0.235\textwidth]{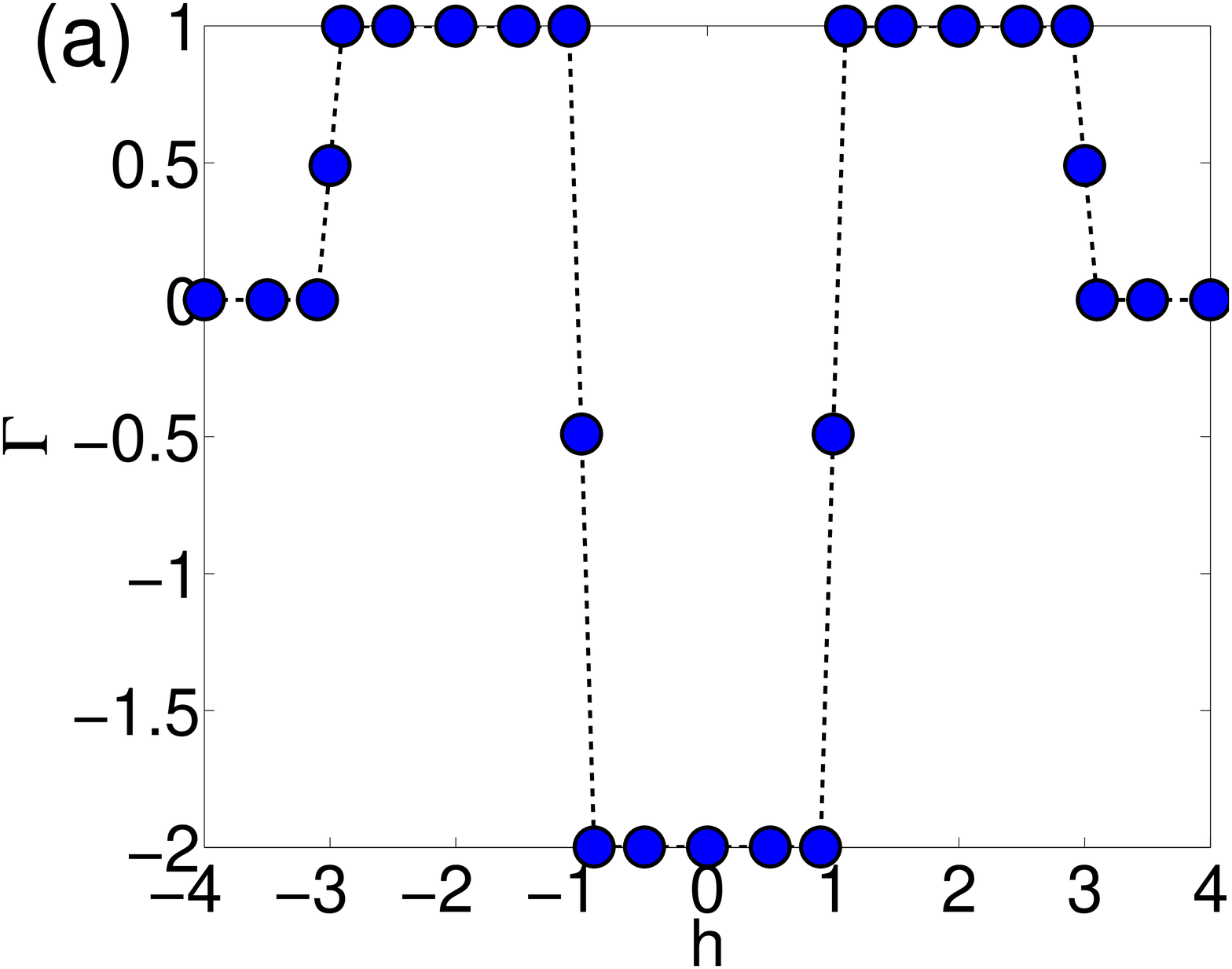}
\includegraphics[trim=0.3cm 0.2cm 2.7cm 1cm, clip,width=0.235\textwidth]{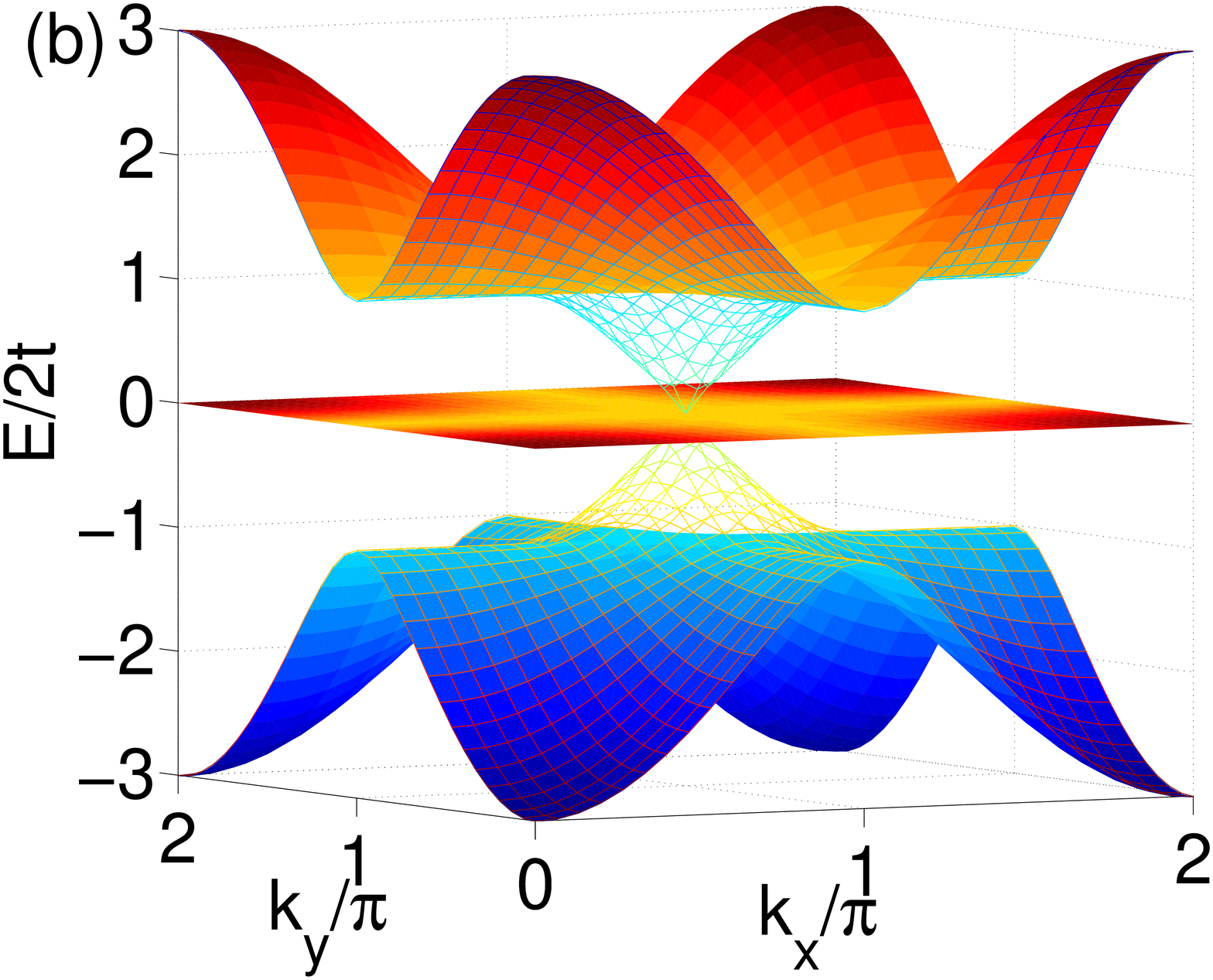} \\
\vspace{-0.6cm}\flushleft (c)\vspace{-.1cm}\\
\includegraphics[trim=9cm 1cm 12cm 1cm, clip,width=0.5\textwidth]{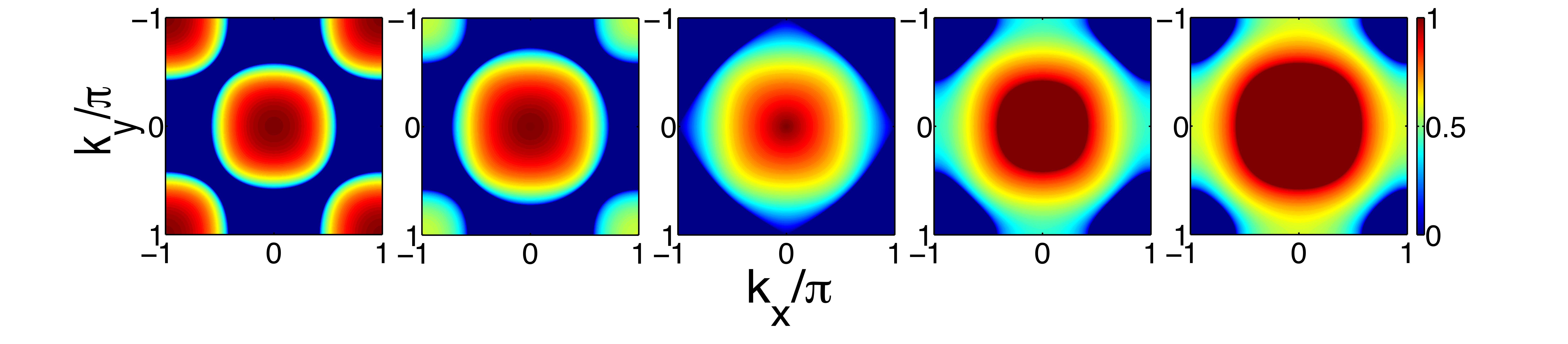}
\caption{(Color online) (a) The topological index $\Gamma$ as a function of the parameter $h$. (b) Energy dispersion for three bulk bands (surface plot) and surface states (mesh plot) at the boundary along the $z$ direction for $h=2$. (c) Quasimomentum distribution $\rho_{\text{cry}}(\mathbf{k})$ for various $h=0,0.5,1,1.5,2$ at a fixed chemical potential $\protect\mu/2t=-2$ \cite{OpticalLatCTI:Sup}. One hundreds layers are taken along the $z$ direction with open boundaries in (b) and (c).}
\label{fig:Index}
\end{figure}

We now proceed to discuss detection methods to probe the exotic phases of
the realized Hamiltonian. The topological index $\Gamma$ defined in Eq.\ \ref{Eqn:Index}
is shown in Fig.\ \ref{fig:Index}(a) under different values of $h$. The system is topologically nontrivial for $|h|<3$, and $\Gamma $ changes at $|h|=1,3$, indicating a topological quantum phase transition. We calculate the band structure numerically for a homogeneous system by keeping $x$ and $y$ directions in momentum space and $z$ direction in real space with open boundaries. Figure \ref{fig:Index}(b) shows the result, revealing the macroscopic flat band as well as the surface states with Dirac cones. Experimentally, the band structure can be probed by mapping out the crystal quasimomentum distribution $\rho_{\text{cry}}(\mathbf{k})$. By abruptly turning off the lattice potential, one could measure the momentum distribution $\rho(\mathbf{k})$, and the quasimomentum can then be extracted as $\rho_{\text{cry}}(\mathbf{k})=\rho(\mathbf{k})/|w(\mathbf{k})|^{2}$, where $w(\mathbf{k})$ is the Fourier transform of the Wannier function $w(\mathbf{r})$ \cite{Spielman2007Mott,*Kashurnikov2002Revealing}. Here, we numerically calculate the crystal quasimomentum distribution, which can be used to track the topological phase transition [Fig.\ \ref{fig:Index}(c)]. At a fixed chemical potential, as one varies $h$ from $0$ to $2$, the quasimomentum distribution reshapes accordingly when the bulk gap closes and reopens and the number of surface Dirac cones changes from $2$ to $1$, indicated by a change of topology of the Fermi surface \cite{OpticalLatCTI:Sup}.

Bragg spectroscopy is a complementary detection method to reveal the Dirac
cone structure \cite{stamper1999excitation,zhu2007simulation}. One could
shine two laser beams at a certain angle to induce a Raman transition from
an occupied spin state to another hyperfine level and focus them near the
surface of the 3D atomic gas. The atomic transition rate can be measured,
which is peaked when the momentum and energy conservation conditions are
satisfied. By scanning the Raman frequency difference, one can map out the
surface energy-momentum dispersion relation \cite{zhu2007simulation}. 
The surface Dirac cones, with their characteristic linear dispersion, can
therefore be probed through Bragg spectroscopy.

So far, we considered a homogeneous system under a box-type trap at zero temperature. In a realistic experiment, finite temperature and a weak confining harmonic trap may introduce noise. To include these effects, an important element to consider is the size of the bulk gap.  In our parameter regime, the minimum band gap from the top or bottom bulk band to the middle flat band is $2t=(2\pi \hbar) \times 80\,$Hz at $h=2$ [Fig.\ \ref{fig:Index}(b)], which corresponds to a temperature around $4 \,$nK. Direct cooling to subnanokelvin temperature is challenging but has been attained experimentally \cite{Leanhardt12092003, *Medley2011Spin}. Parametric cooling based on adiabatic preparation can be used to further reduce the effective temperature of the system. With a band gap considerably larger than the probing Raman Rabi frequency, bulk contribution to the Bragg spectroscopy is negligible. In the following, we include the effect of a weak harmonic trap via the local density approximation (LDA) and consider the finite temperature effects to be minimal.

\begin{figure}[t]
\includegraphics[trim=0.55cm 1.1cm 2.3cm 1.3cm, clip,width=0.245\textwidth]{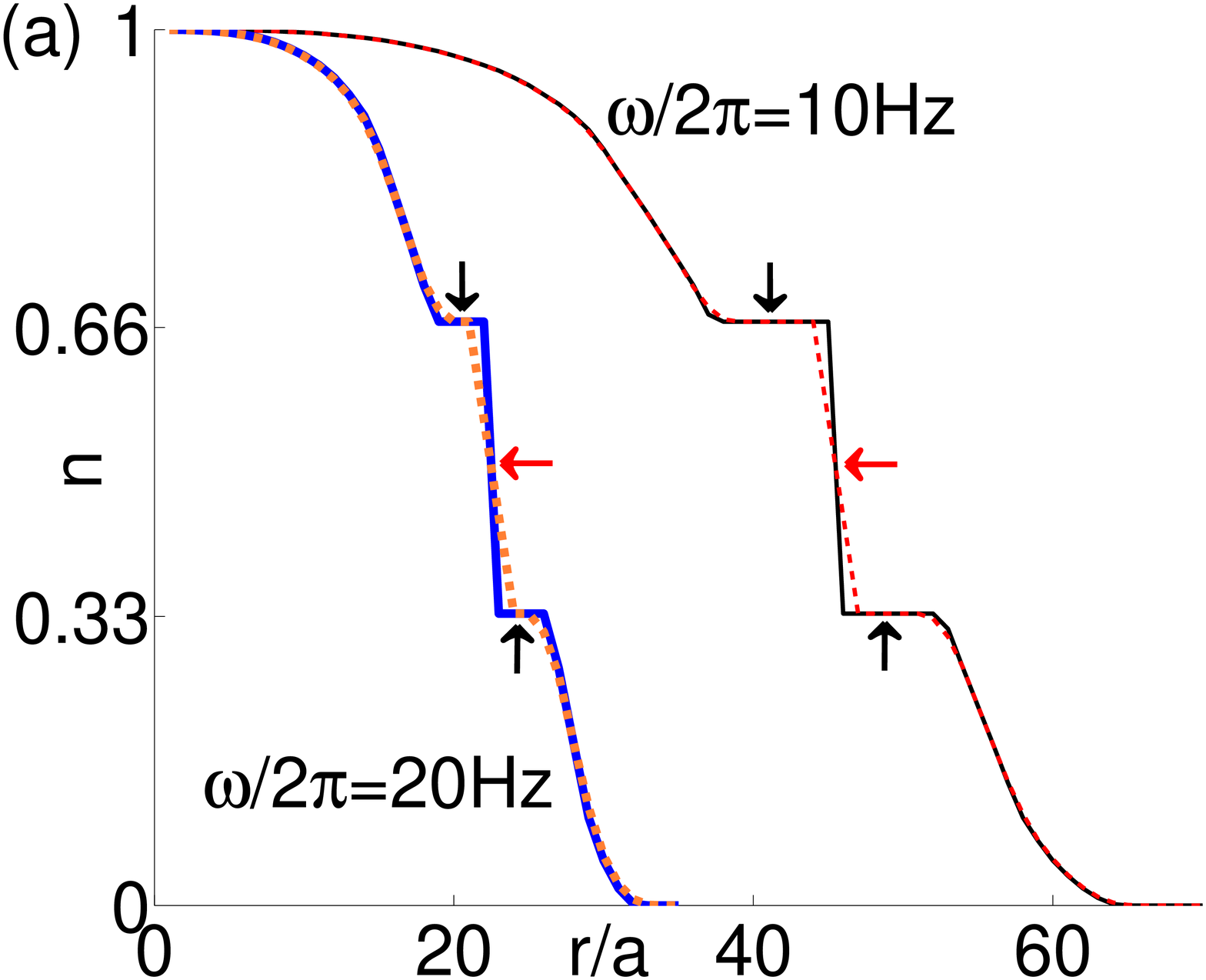}
\hspace{-.4cm}\includegraphics[trim=.55cm 1.1cm 2.3cm 1.3cm, clip,width=0.245\textwidth]{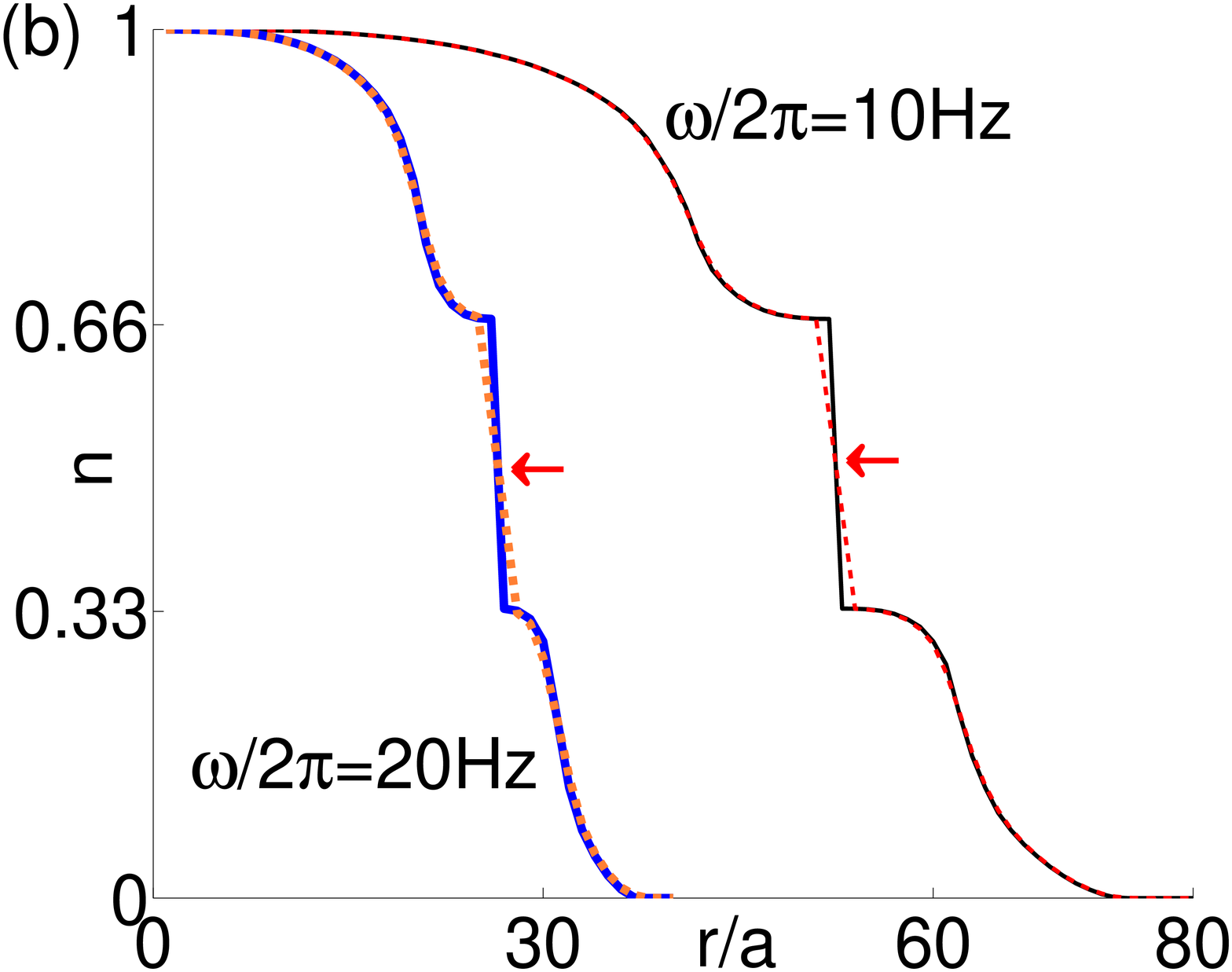}
\caption{The atomic density profile $n$ as a function of the radial distance
$r$ under the LDA. (a) $h=0, \protect\mu_{0}/2t=3$. (b) $h=1,\protect\mu_{0}/2t=4$. $^{40}$K is used and $t/\hbar$ is taken to be $2\pi \times 40\,$Hz.}
\label{fig:LDA}
\end{figure}

The characteristic flat band can be detected through measurement of the atomic density profile under the global harmonic trap \cite{bloch2005ultracold,Schneider05122008}.  Under the LDA, the local chemical potential of the system is $\mu (r)=\mu_{0}-m\omega ^{2}r^{2}/2$, where $\mu _{0}$ denotes the chemical potential at the center of a spherically harmonic trap with the potential $V(r)=m\omega^{2}r^{2}/2$. The local atomic density $n(r)$ is uniquely determined by $\mu(r)$, and $\mu _{0}$ is specified by the total atom number $N$ through $\int n(r)4\pi r^{2}\,dr=N$. The atomic density profile $n(r)$, which can be measured \emph{in situ} in experiments \cite{Schneider05122008}, is calculated and shown in Fig.\  \ref{fig:LDA}. A steep fall or rise in $n(r)$ is a clear signature of a macroscopic flat band (horizontal arrows in Fig.\ \ref{fig:LDA}). The plateaus at $1/3$ and $2/3$ fillings [vertical arrows in Fig.\ \ref{fig:LDA}(a)] reveal the corresponding band gap. At $h=1$, the plateaus vanish [Fig.\ \ref{fig:LDA}(b)]. The disappearance of the plateaus at this point indicates the phase transition where the band gap closes. In experiments, due to the
finite spatial resolution, the detected signal may correspond to a locally
averaged $n(r)$. The dashed lines show the local average density $\bar{n}%
_{i}=\sum_{j=-1}^{1}n_{i+j}/3$, averaged over a spherical shell of $3$
lattice sites. One can see that major features associated with the band
gap and the flat band remain clearly visible even when the signal is blurred by the
local spatial averaging.

In summary, we have proposed an experimental scheme to realize and probe a 3D chiral
TI with a zero-energy flat band. The experimental realization of this
model will mark an important advance in the ultracold atom simulation of
topological phases.

\begin{acknowledgments}
We thank R.\ Ma, S.-L.\ Zhu, C.-J.\ Wu, K.\ Sun, and G.\ Ortiz for
discussions. This work was supported by the NBRPC (973 Program) No.\ 2011CBA00300 (No.\ 2011CBA00302), the IARPA MUSIQC program, the ARO, and the AFOSR MURI program.
\end{acknowledgments}


%

\onecolumngrid
\appendix
\clearpage
\section*{Supplemental Material: Probe of Three-Dimensional Chiral Topological Insulators in an Optical Lattice}

\begin{quote}
In this supplemental material, we provide more details on the realization scheme of the chiral topological insulator Hamiltonian. Details on the parameter estimation with Wannier functions and additional density of states plots are included.  
\end{quote}

\section{Realization of the effective Hamiltonian}

\begin{figure}[h]
\hspace{-7cm}(a)  \hspace{6.5cm} (b) \\
\vspace{-.2cm}\includegraphics[trim=0cm 0cm 0cm 0cm, clip, width=0.3\textwidth]{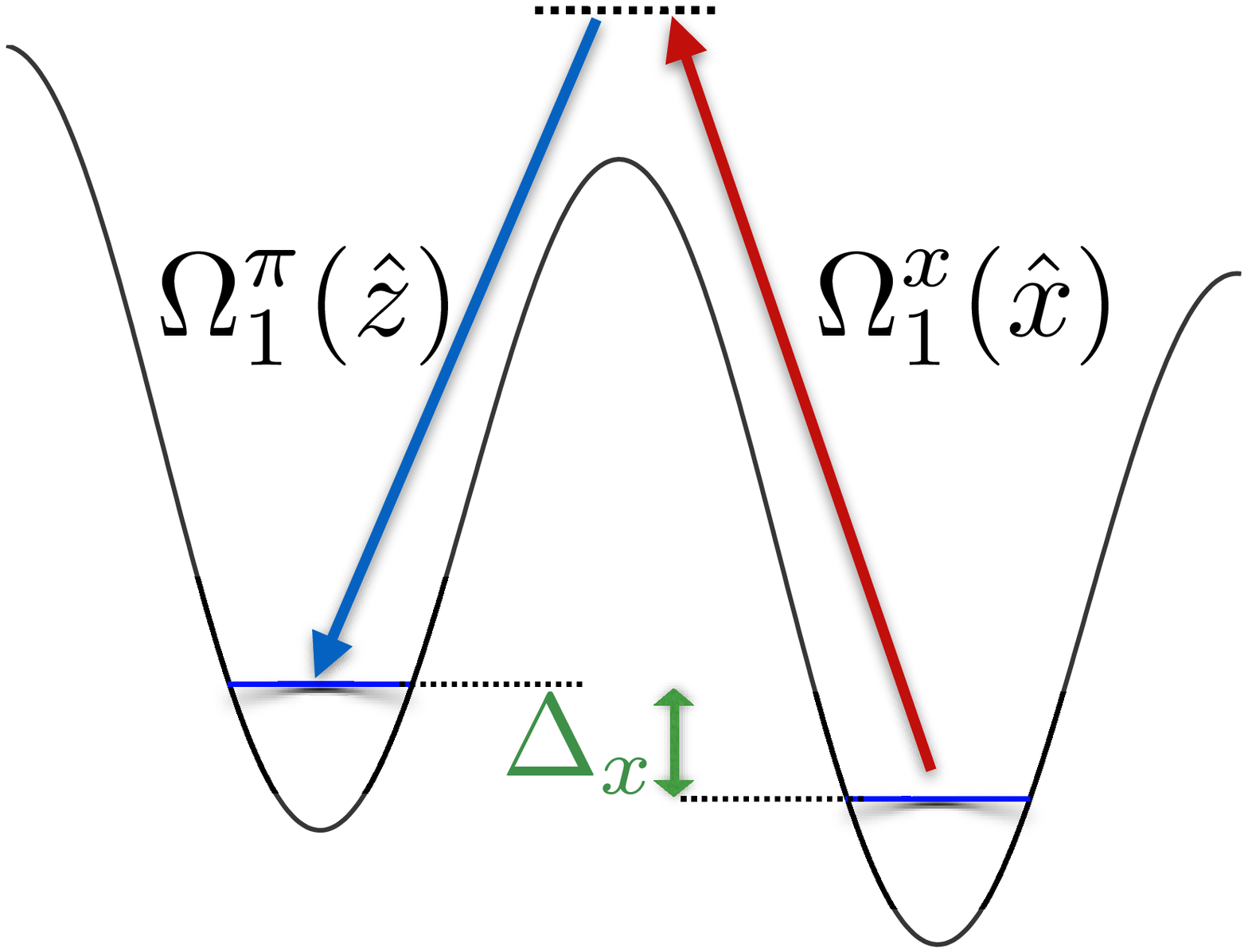} \hspace{1.5cm}
\includegraphics[trim=0cm 0cm 0cm 0cm, clip, width=0.35\textwidth]{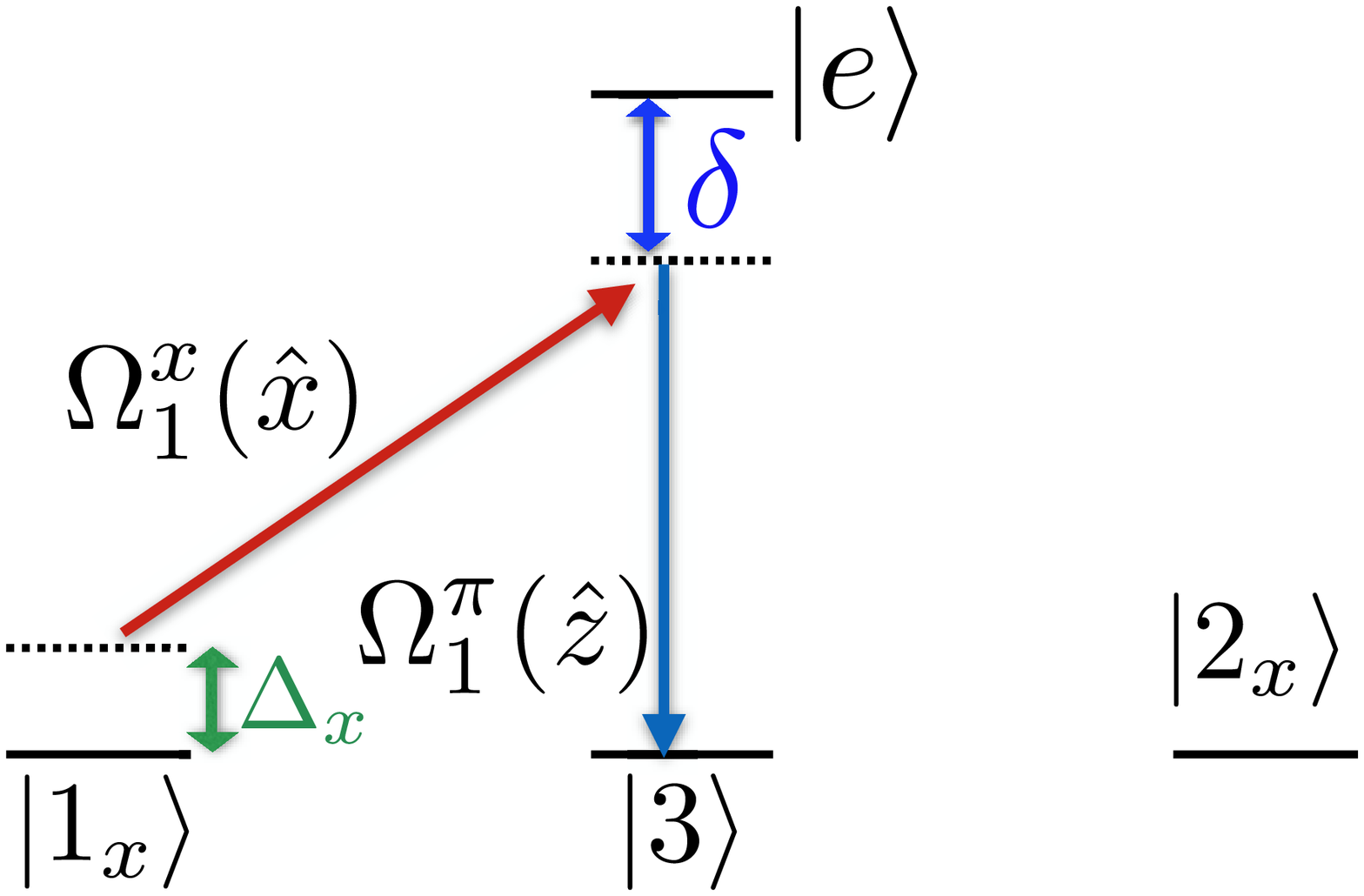}
\caption{(a) A linear tilt $\Delta_{x}$ per site in the lattice along $x$-direction. (b) Two Raman beams $\Omega _{1}^{x}$ and $\Omega _{1}^{\pi}$ used to produce the tunneling in $H^{(1)}_{\mathbf{rx}}$. The unit vectors in brackets show the polarization direction of the corresponding beam. For the complete optical lattice setup, refer to Fig.\ 1 in the main text.}
\label{FigSup:Lattice}
\end{figure}

Our main result is to use Raman-assisted tunneling to realize the effective Hamiltonian of the chiral topological insulator given by
\begin{align}
H&= t \sum_{\mathbf{r}}\left[ \left( 2ihc_{3,\mathbf{r}%
}^{\dag }c_{2,\mathbf{r}}+\text{H.c.}\right) +H_{\mathbf{rx}}+H_{\mathbf{ry}%
}+H_{\mathbf{rz}}\right] \notag, \\
H_{\mathbf{rx}}& =ic_{3,\mathbf{r-x}}^{\dag }(c_{1,\mathbf{r}}+c_{2,\mathbf{%
r}})-ic_{3,\mathbf{r+x}}^{\dag }(c_{1,\mathbf{r}}-c_{2,\mathbf{r}})+%
\text{H.c.,}   \\
H_{\mathbf{ry}}& =-c_{3,\mathbf{r-y}}^{\dag }(c_{1,\mathbf{r}}-ic_{2,\mathbf{%
r}})+c_{3,\mathbf{r+y}}^{\dag }(c_{1,\mathbf{r}}+ic_{2,\mathbf{r}})+%
\text{H.c.,}  \notag \\
H_{\mathbf{rz}}& =2ic_{3,\mathbf{r-z}}^{\dag }c_{2,\mathbf{r}}+\text{H.c.} \notag
\label{EqnSup:Ham2}
\end{align}%
In the following, we provide some complementary details on the realization scheme. The major difficulty is to realize the spin-transferring hopping terms $H_{\mathbf{rx}%
},H_{\mathbf{ry}},H_{\mathbf{rz}}$ along each direction. Let us focus on a single term first, $H^{(1)}_{\mathbf{rx}}=ic_{3,\mathbf{r-x}}^{\dag }(c_{1,\mathbf{r}}+c_{2,\mathbf{%
r}})$. This corresponds to an atom in the spin state $\ket{1_x}=\left(\ket{1}+\ket{2}\right)/\sqrt{2}$ at site $\mathbf{r}$ hopping to site $\mathbf{r-x}$ while changing the spin state to $\ket{3}$ with hopping strength $i \sqrt{2}$. Diagrammatically, it can be visualized as
\begin{equation}
ic_{3,\mathbf{r-x}}^{\dag }(c_{1,\mathbf{r}}+c_{2,\mathbf{%
r}}) \qquad \Longleftrightarrow \qquad x\text{-}\text{direction:}\quad \ket{3}\overset{i\sqrt{2}}{\curvearrowleft}\ket{1_x}\overset{\times }{\curvearrowright }\;
\end{equation}
where $\overset{i\sqrt{2}}{\curvearrowleft}$ means hopping along that direction with strength $i\sqrt{2}$ and $\overset{\times}{\curvearrowright}$ indicates hopping is forbidden. This hopping term can be effected by two Raman beams $\Omega _{1}^{x}=i\sqrt{2}\Omega _{0}e^{ikz}$ and $\Omega _{1}^{\pi}=\Omega _{0}e^{ikx}$ as shown in Fig.\ \ref{FigSup:Lattice}.

The large single-photon detuning $\delta$ ensures that the population of the excited state, estimated by $|\Omega_{0}/\delta|^{2}$, is negligible. The two-photon detuning $\Delta_{x}$ matches the linear energy shift of the lattice per site, so that it only allows $\ket{1_{x}}$ to hop to the left, and the other direction is forbidden by an energy mismatch $2\Delta_{x}$. The addressing of spin states is done by polarization selection rule. The original spin basis $\ket{1},\ket{3}, \ket{2}$ differ in the magnetic quantum number $m$ by one successively. So a $\pi$-polarized beam $\Omega_{1}^{\pi}$ excites the state $\ket{3}$ and a linear $\hat{x}$-polarized beam $\Omega_{1}^{x}$ excites the superposition state $\ket{1_x}=\left(\ket{1}+\ket{2}\right)/\sqrt{2}$ since $\hat{x} \sim (\sigma^{+}+\sigma^{-})$. So together, these two beams induce a Raman-assisted hopping between $\ket{1_{x}}$ and $\ket{3}$. The hopping amplitude and phase are controlled by the corresponding Raman beam amplitude and phase. In addition, the wave-vector difference of two beams $\delta \mathbf{k}$ ($(-k, 0, k)$ in this case) has to have a component along the hopping direction ($x$-direction) to ensure the hopping strength is non-vanishing. 

All the other hopping terms in the Hamiltonian are realized in a similar manner. For example, consider the term $c_{3,\mathbf{r+y}}^{\dag }(c_{1,\mathbf{r}}+ic_{2,\mathbf{r}})$, which can be realized by $\Omega _{2}^{y}=\sqrt{2}\Omega _{0}e^{ikz}$ and $\Omega _{2}^{\pi}=\Omega _{0}e^{iky}$, polarized along $(\hat{x}+\hat{y})$-direction and $\hat{z}$-direction respectively (see Fig.\ 1 in the main text). Since $(\hat{x}+\hat{y}) \sim (\sigma^{+}+i\sigma^{-})$, it couples the state $\ket{2_y}=(\ket{1}+i\ket{2})/\sqrt{2}$ and $\ket{3}$. A wave-vector difference $\delta \mathbf{k} = (0,-k,k)$ and a two-photon energy detuning $\Delta_{y}$ guarantee the desired hopping along $y$-direction.

With a number of laser beams required to realize the full Hamiltonian, it is important to check that undesired tunneling terms are forbidden. To that end, we require a different
linear energy shift per site $\Delta _{x,y,z}$ along the $\left(
x,y,z\right) $-direction. The ratio between $\Delta _{x,}\Delta _{y,}\Delta_{z}$ can be adjusted by setting the direction of the gradient field to be in a specific angle with respect to the three axes of the optical lattice. In particular, we set $\Delta _{x}:\Delta _{y}:\Delta _{z}=1:2:3$. The energy difference is lower bounded by $\Delta_{x}$. So if we select a parameter regime such that the Raman-assisted hopping rate $t$ satisfies $t\ll \Delta _{x}$, then the hopping along the $z$ direction induced by $\Omega _{1}^{x}$ and $\Omega _{1}^{\pi }$, for instance, have negligible effects because of the large detuning. Other undesired couplings between different beams are disallowed because the wave-vector difference $\delta \mathbf{k}$ may not have the component along a certain direction to induce a hopping along that direction. For example,   $\Omega _{1}^{x}=i\sqrt{2}\Omega _{0}e^{ikz}$ and $\Omega _{2}^{\pi}=\Omega _{0}e^{iky}$ will not induce a hopping along $x$-direction as $\delta \mathbf{k}$ does not include a component along $x$-direction. Moreover, the Raman beams $\Omega _{1,2}^{x,y,z}$ and $\Omega _{1,2}^{\pi}$ may induce some on-site spin transferring terms, which can be compensated with some radio-frequency fields. 

\section{Wannier-(Stark) Function Estimation}

\begin{figure}[b]
\includegraphics[trim=0cm .3cm 4.3cm 1.4cm, clip,width=0.8\textwidth]{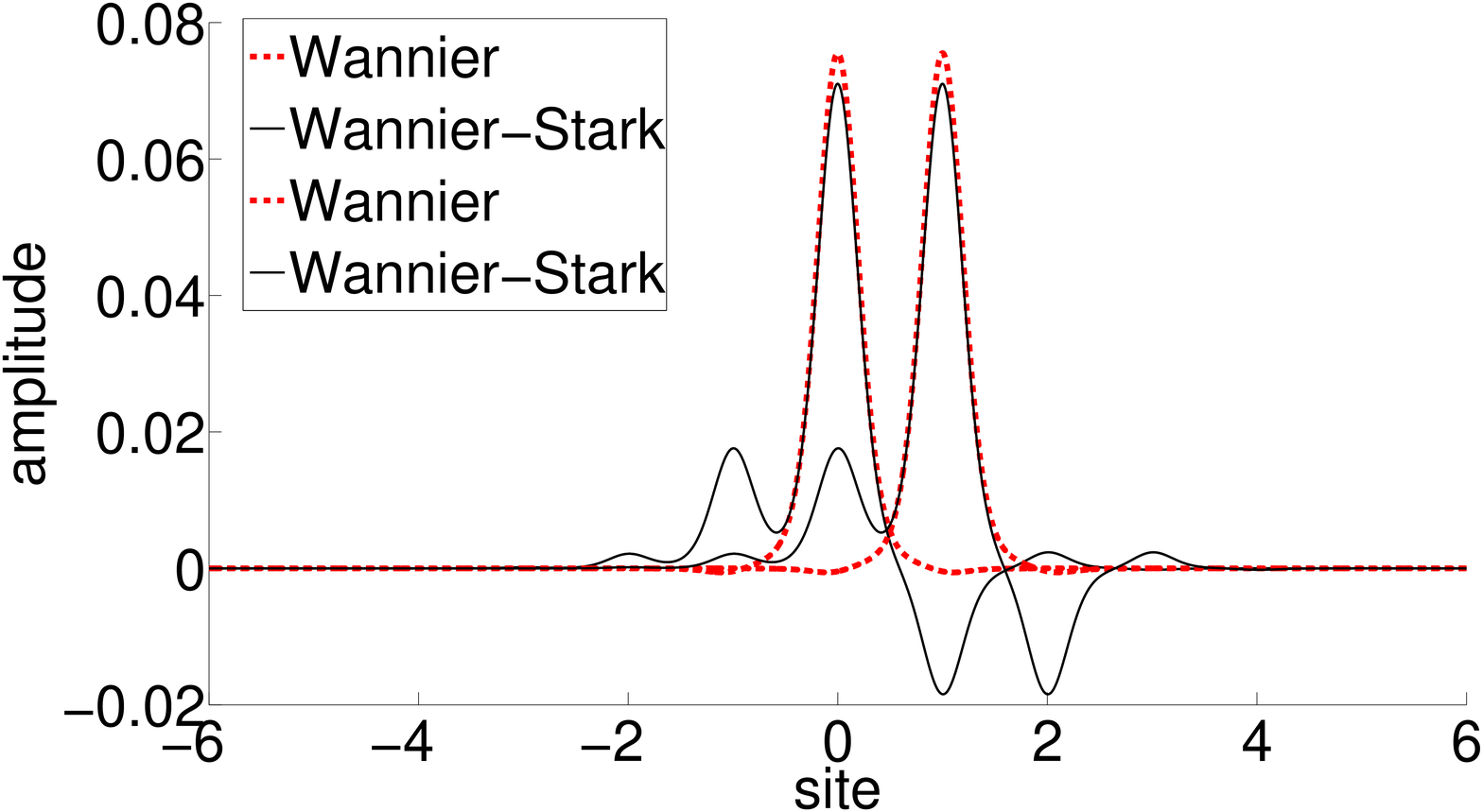}
\caption{Wannier functions and Wannier-Stark functions centered at site $0$ and site $1$, with $V_{0} \approx 2.3 E_{r}, a=2\pi/k = 764\,$nm, and linear tilt $\Delta_{x}/2\pi \approx 200\,$Hz for $^{40}K$ atoms. Note that the apparent shallow potential is due to the larger separation between two sites. We take $a=\lambda$ instead of $a=\lambda/2$ in the typical case (see main text).}
\label{FigSup:Wannier}
\end{figure}

In the second quantization representation with the Wannier function basis, the natural tunneling rate can be written as
\begin{equation}
t_{0} =\int d^{3}\mathbf{r}^{\prime}
\bar{w}^{\ast }(\mathbf{r}^{\prime }-\mathbf{r_{i}})
\left[ -\dfrac{\hbar^{2}}{2m} \nabla^{2} + V_{0}(\mathbf{r^{\prime}})
\right]
 \bar{w}(\mathbf{r}^{\prime }-\mathbf{r_{j}}),
\end{equation}
where $ \bar{w}(\mathbf{r}^{\prime }-\mathbf{r_{j}})$ is the Wannier function centered at site $\mathbf{r_{j}}$ and $V_{0}(\mathbf{r^{\prime}})$ is the lattice depth at site $\mathbf{r_{j}}$ (We use $\bar{w}(\mathbf{r})$ to denote the Wannier function and $w(\mathbf{r})$ to denote the Wannier-Stark function to be notationally consistent with the main text). With a linear tilt in the optical lattice, translational symmetry is broken and Wannier functions are no longer the proper descriptions of the localized states. Instead, a simple modification with Wannier-Stark functions $w(\mathbf{r})$ will be sufficient \cite{gluck2002wannier,miyake2013probing}: 
\begin{equation}
w_{i}(\mathbf{r^{\prime}}-\mathbf{r}_{l}) = \sum_{m} J_{m-l} 
\left( \dfrac{2t_{0}}{\Delta_{i}} \right) \bar{w}(\mathbf{r^{\prime}}-\mathbf{r_{m}}),
\end{equation}
where $i=x,y,z$, and $w_{i}(\mathbf{r})$ is the Wannier-Stark function, and $\Delta_{i}$ is the linear tilt per site along $i$ direction. $J_{m-l} (x)$ are the bessel functions of the first kind. Fig.\ \ref{FigSup:Wannier} shows the Wannier functions and Wannier-Stark functions with $V_{0} \approx 2.3 E_{r}$. They have close overlaps on the center site, but may differ significantly on neighboring sites. Calculations with the Wannier functions or the Wannier-Stark functions produce the same natural tunneling $t_{0}/ \hbar \approx 2\pi \times 50\,$Hz. The tunneling rate with Raman-assisted hopping can be written as an integral of Wannier-Stark functions (as discussed in the main text):
\begin{equation}
t_{\mathbf{r,m}}=\dfrac{\Omega _{\beta \mathbf{m}}^{\ast }\Omega _{\alpha
\mathbf{m}}}{\delta }\int d^{3}\mathbf{r}^{\prime }w^{\ast }(\mathbf{r}%
^{\prime }-\mathbf{r-m})e^{i\delta \mathbf{k}\cdot \mathbf{r}^{\prime }}w(%
\mathbf{r}^{\prime }-\mathbf{r}).
\label{EqnSup:hopping}
\end{equation}
Factorizing the Wannier-Stark functions into each direction, $w(\mathbf{r}^{\prime
})=w(x^{\prime })w(y^{\prime })w(z^{\prime })$ and calculating them along each direction, we can numerically compute the overlap integral 
\begin{equation*}
\beta \equiv \int dxw^{\ast
}(x+a)e^{-ikx}w(x)\int dyw^{\ast }(y)w(y)\int dzw^{\ast }(z)e^{ikz}w(z). 
\end{equation*}
With the parameters given in Fig.\ \ref{FigSup:Wannier}, we have $\beta \approx 0.34$. For Raman beams with $\Omega_{0}/2\pi \approx 15\,$MHz and the single-photon detuning $\delta/2\pi \approx 1.7\,$THz, we have $\Omega_{\mathbf{R}} = |\Omega_{0}|^{2}/
\delta \approx  2\pi \times 120\,$Hz, and the Raman-assisted hopping rate $t/\hbar \approx 2\pi \times 40\,$Hz. It is worthwhile to point out that the expression given in equation \eqref{EqnSup:hopping} is only valid in the perturbative limit when $t \lesssim t_{0}$. When the Rabi frequency becomes stronger, the Raman-assisted tunneling rate eventually saturates. A more accurate expression may be obtained in the nonperturbative limit with a more accurate analysis \cite{miyake2013probing, Miyake:2013jw}. Nevertheless, these numerical calculations only yield rough estimations to experimental parameters, which may need to be fine-tuned in experiments to produce the best result in a topologically nontrivial phase.

\section{Density of States}
\begin{figure}[h]
\includegraphics[trim=0.4cm 0.4cm 2.4cm 2cm, clip, width=\textwidth]{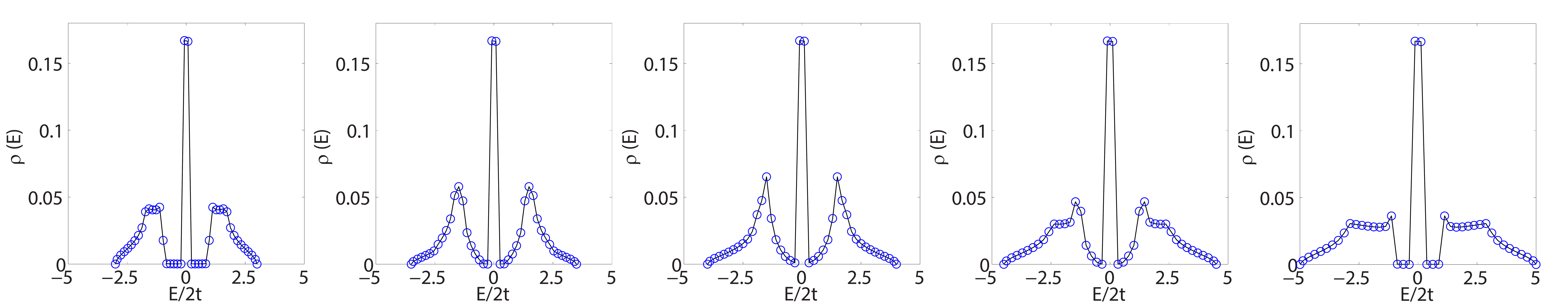}
\caption{Density of states $\rho(E)$ with respect to the energy $E$ for various values of $h$ ($h=0,0.5,1,1.5,2$ from left to right). The middle flat band is clearly visible at zero energy. The bulk band gaps are $2t, t/2, 0, t/2, 2t$ respectively.}
\label{FigSup:Phase}
\end{figure}

In Fig.\ 2(c) of the main text, we plotted the momentum distribution of atoms $\rho_{\text{cry}}(\mathbf{k})$ for various $h$ at a fixed chemical potential $\mu/2t=-2$. It is useful to include the density of states $\rho(E)$ for various values of the parameter $h$. In Fig.\ \ref{FigSup:Phase} here, we show the density of states plots. The macroscopic zero-energy flat band is prominent in each plot. The band gap is also clearly visible for $h=0,2$ (less visible for $h=0.5,1.5$). In Fig.\ 2(c) of the main text, the figures correspond to a filling up to $\mu/2t=-2$. A change of Fermi surface topology can be observed in those figures. 


%

\end{document}